\documentclass{article}

\usepackage[english]{babel}
\usepackage[letterpaper,top=2cm,bottom=2cm,left=3cm,right=3cm,marginparwidth=1.75cm]{geometry}

\usepackage{amsmath}
\usepackage{graphicx}
\usepackage[colorlinks=true, allcolors=blue]{hyperref}
\usepackage{booktabs}
\usepackage{tablefootnote}
\usepackage{multirow}
\usepackage{subcaption}

\title{Dstack: A Zero Trust Framework for Confidential Containers}

\author{
Shunfan Zhou\\
Phala Network\\
\texttt{shelvenzhou@phala.network}
\and
Kevin Wang\\
Phala Network\\
\texttt{kvinwang@phala.network}
\and
Hang Yin\\
Phala Network\\
\texttt{hangyin@phala.network}
}

\begin{document}
\maketitle

\begin{abstract}
    Web3 applications require execution platforms that maintain confidentiality and integrity without relying on centralized trust authorities. While Trusted Execution Environments (TEEs) offer promising capabilities for confidential computing, current implementations face significant limitations when applied to Web3 contexts, particularly in security reliability, censorship resistance, and vendor independence.

    This paper presents \textit{dstack}, a comprehensive framework that transforms raw TEE technology into a true Zero Trust platform. We introduce three key innovations: (1) \textit{Portable Confidential Containers} that enable seamless workload migration across heterogeneous TEE environments while maintaining security guarantees, (2) \textit{Decentralized Code Management} that leverages smart contracts for transparent governance of TEE applications, and (3) \textit{Verifiable Domain Management} that ensures secure and verifiable application identity without centralized authorities.

    These innovations are implemented through three core components: dstack-OS, dstack-KMS, and dstack-Gateway. Together, they demonstrate how to achieve both the performance advantages of VM-level TEE solutions and the trustless guarantees required by Web3 applications. Our evaluation shows that dstack provides comprehensive security guarantees while maintaining practical usability for real-world applications.
\end{abstract}

\section{Introduction}
\label{sec:introduction}

The Web3 ecosystem represents a paradigm shift in digital trust models, fundamentally transforming the relationship between users and application developers.
Unlike traditional systems, Web3 empowers users with unprecedented control over their data through the principle of ``Code is Law''—once deployed, code operates autonomously and independently of external control.
This transformation demands robust execution platforms that maintain confidentiality and integrity without relying on centralized trust authorities.

\begin{figure}[ht]
    \centering
    \includegraphics[width=0.8\textwidth]{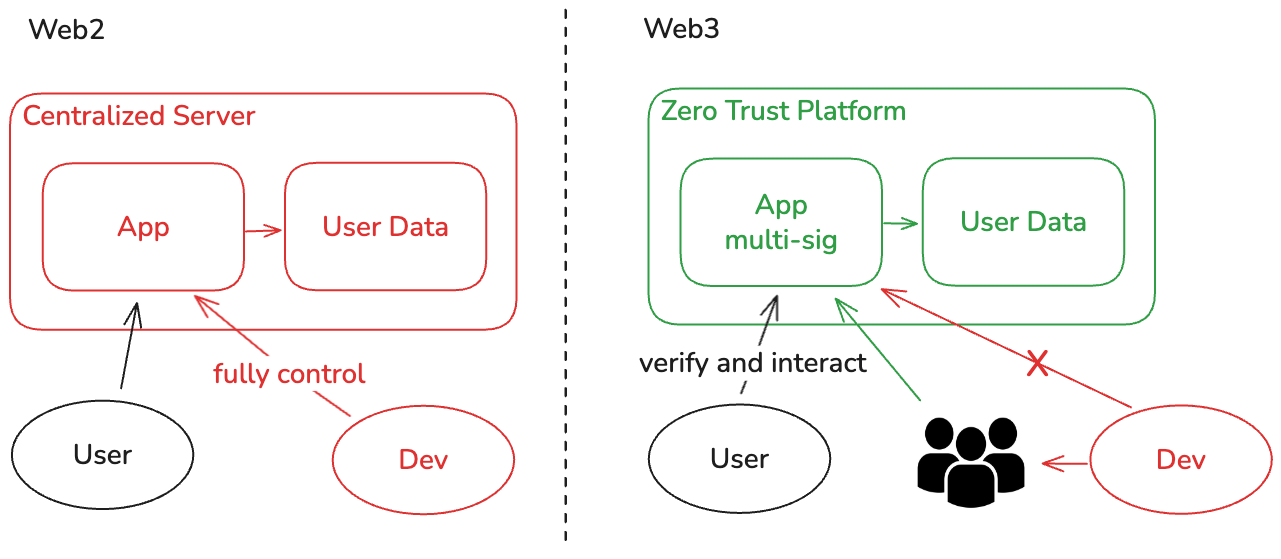}
    \caption{\small The evolution of trust models from Web2 to Web3, highlighting the shift from centralized control to decentralized governance and execution.}
    \label{fig:comparison}
\end{figure}

Figure~\ref{fig:comparison} illustrates this fundamental shift from Web2 to Web3 architectures.
In Web2 systems, developers retain complete control through proprietary applications and centralized servers, with administrative privileges that enable unrestricted access to user data.
In contrast, Web3 introduces smart contracts as the first significant step toward true trustlessness by executing programs on decentralized platforms (blockchains).
Once deployed, administrators cannot alter program execution, and lifecycle management (such as code updates) is governed by predefined agreements through democratic mechanisms including multi-signature wallets and Decentralized Autonomous Organizations (DAOs).
This emerging trust model extends beyond basic applications to mission-critical scenarios, such as AI model training where data providers and model developers may have competing interests, and autonomous AI agents designed to operate independently of human intervention.
In these contexts, Zero Trust platforms are essential for hosting applications and enforcing verification mechanisms that appropriately constrain developer authority.

A Zero Trust platform must enforce four key principles:

\begin{itemize}
    \item \textbf{Code is Law}: This principle manifests in two dimensions: (1) application logic, once deployed, cannot be changed unexpectedly; and (2) code lifecycle management—including deployment, upgrades, and deletion—follows predefined governance rules.
    \item \textbf{Censorship Resistance}: User data must remain beyond the control of any single entity, with data availability protected against denial-of-service attacks and other forms of censorship.
    \item \textbf{Full Chain of Trust}: Users must be able to verify every aspect of an application, including network configuration, application identity and code logic, underlying hardware, and execution environment.
    \item \textbf{Assume Breach}: The platform must operate under the assumption that compromises will occur, implementing mechanisms for damage containment, rapid recovery, and minimizing data exposure during security incidents.
\end{itemize}

Confidential Computing, particularly through Trusted Execution Environments (TEEs), has emerged as a promising foundation for extending blockchain capabilities to support complex computational tasks with confidentiality guarantees.
Recent advances in VM-level TEE solutions, such as Intel TDX~\cite{li2019intel} and AMD SEV~\cite{kaplan2016amd}, have demonstrated substantial performance improvements and developer-friendly interfaces, making them compelling candidates for Web3 development.
The integration of GPUs within TEEs~\cite{tramer2018slalom,hunt2018chiron} now enables verifiable execution of both Large Language Models and AI agents, providing cryptographic guarantees of computational integrity and privacy.

However, a significant gap remains between raw TEE technology and the requirements of a true Zero Trust platform.
Current TEE implementations face several critical limitations when applied to Web3 contexts, which directly undermine the Zero Trust principles outlined above:

\begin{itemize}
    \item \textbf{Security Reliability}: Recent vulnerabilities in TEE systems have raised concerns about their effectiveness as trust anchors. Persistent side-channel attacks~\cite{van2018foreshadow,murdock2020plundervolt,van2020lvi} and micro-architectural flaws undermine confidence in TEE security, especially for high-value Web3 applications where a single compromise could result in catastrophic data exposure, thereby violating the ``Assume Breach'' principle.
    \item \textbf{Censorship Vulnerability}: Conventional TEE encryption schemes bind keys to specific hardware instances, creating single points of failure. This approach not only threatens data availability but also introduces vendor dependencies, as centralized hardware manufacturers can become vectors for censorship or control, contradicting the ``Censorship Resistance'' principle.
    \item \textbf{Incomplete Verifiability}: While TEEs provide Remote Attestation to verify application and hardware identity, they do not deliver the comprehensive chain of trust required in Zero Trust environments. Users lack guarantees that programs will process their data as expected and cannot be modified without proper authorization, falling short of the ``Full Chain of Trust'' requirement.
    \item \textbf{Unrestricted Application Lifecycle Control}: Existing TEE platforms lack robust mechanisms to prevent developers from updating deployed applications from benign to malicious versions. Without decentralized and transparent code management, users cannot be assured that application upgrades or modifications follow agreed-upon governance, leaving the system vulnerable to insider threats and unauthorized changes, again undermining ``Code is Law''.
\end{itemize}

To bridge this gap, we present \textit{dstack}, a comprehensive framework for confidential containers that introduces three key innovations. Each innovation directly addresses the limitations outlined above while reinforcing Zero Trust principles:

\begin{itemize}
    \item \textbf{Portable Confidential Container}: Our architecture enables seamless migration of confidential workloads across different TEE instances and hardware vendors, significantly reducing vendor lock-in risks while maintaining robust security guarantees. This innovation addresses censorship vulnerability and vendor dependency by providing hardware abstraction and state continuity across heterogeneous TEE environments.
    \item \textbf{Decentralized Code Management}: We implement a comprehensive governance framework that leverages smart contracts for transparent and decentralized management of TEE applications. This system ensures verifiable deployment and upgrade processes, enforcing the ``Code is Law'' principle by binding application lifecycles to on-chain governance decisions. This innovation addresses incomplete verifiability and uncontrolled program management by creating an immutable audit trail of application changes.
    \item \textbf{Verifiable Domain Management}: Our novel approach to certificate management ensures that confidential containers can exclusively control domains and provide native HTTPS support to TEE applications without relying on centralized authorities. Acting as middleware between TEE applications and clients, this system enables users to establish end-to-end encrypted channels with TEE applications without requiring modifications to client software.
\end{itemize}

These innovations are realized through a cohesive system architecture comprising three core components:

\begin{itemize}
    \item \textbf{dstack-OS}: A hardware abstraction layer with a minimized operating system image that eliminates differences in underlying TEE hardware while reducing the attack surface. This component provides a consistent, secure runtime environment across diverse TEE implementations, supporting both portability and reliability.
    \item \textbf{dstack-KMS}: A blockchain-controlled key management service that replaces hardware-based encryption schemes with an independent service for generating and managing secret keys. This component enables secure data migration across TEE instances and supports key rotation to provide forward and backward data secrecy, thereby mitigating the impact of potential TEE compromises.
    \item \textbf{dstack-Ingress and dstack-Gateway}: Complementary systems that provide TEE-controlled domain management through different approaches—dstack-Ingress enables applications to specify custom domains with minimal integration requirements, while dstack-Gateway offers pre-registered wildcard domains that require no application code changes.
\end{itemize}

By systematically addressing the critical limitations of current TEE implementations, dstack establishes a comprehensive framework that transforms raw TEE technology into a true Zero Trust platform aligned with Web3's foundational principles. Our approach not only enhances the security and reliability of confidential computing but also provides a practical path toward truly decentralized, censorship-resistant infrastructure for next-generation applications.

The remainder of this paper details the technical design of dstack, evaluates its performance and security characteristics, and demonstrates how our innovations collectively create a Zero Trust platform. This platform maintains the performance advantages of VM-level TEE solutions while overcoming their inherent limitations.

\section{Background and Related Work}

\subsection{Fundamentals of Trusted Execution Environments}

Trusted Execution Environments (TEEs) represent a secure area within a processor that guarantees code and data loaded inside are protected with respect to confidentiality and integrity. Modern TEE implementations can be categorized into two main approaches: enclave-based solutions like Intel SGX, and VM-based solutions such as Intel TDX and AMD SEV-SNP. While enclave-based TEEs provide fine-grained isolation at the process level, VM-based solutions offer better compatibility with existing cloud infrastructure and development practices by securing entire virtual machines.

The security guarantees of TEEs are founded on hardware-based isolation and remote attestation mechanisms. Hardware isolation ensures that even privileged software, including the operating system and hypervisor, cannot access the protected memory regions. Remote attestation enables third parties to verify the integrity of the TEE environment and the software running within it, establishing a chain of trust from hardware to application. However, this traditional approach creates hardware dependencies that limit portability and introduce centralization risks.

\subsection{Evolution of Confidential Computing in Cloud Infrastructure}

Recent years have witnessed significant advancement in confidential computing technologies, particularly in cloud environments. Cloud service providers have increasingly adopted TEE solutions to offer confidential computing services~\cite{gcptee,awsnitro,azuretee}, with major platforms like Azure Confidential Computing and Google Cloud Confidential Computing leading the way. These services primarily leverage VM-level TEE solutions due to their superior performance characteristics and compatibility with existing virtualization infrastructure.

However, current implementations face several limitations when applied to Web3 scenarios. Traditional cloud-based confidential computing assumes a centralized trust model where the cloud provider maintains significant control over the infrastructure. This model conflicts with Web3's fundamental requirement for trustless and decentralized operations, creating a need for new approaches that maintain confidentiality while embracing decentralization.

\subsection{Zero Trust Architecture Principles}

Zero Trust Architecture (ZTA) has emerged as a crucial security paradigm that assumes no implicit trust in any component of the system, regardless of its location or ownership. In the context of Web3 infrastructure, ZTA principles must be extended beyond traditional network security to encompass:

\begin{itemize}
\item Continuous verification of both hardware and software components
\item Cryptographic proof of correct execution
\item Decentralized governance and control mechanisms
\item Minimal trust requirements on hardware manufacturers
\end{itemize}

Traditional ZTA implementations rely on centralized policy enforcement points, which contradicts Web3's decentralized nature. Our work addresses this challenge by implementing ZTA principles through blockchain-based governance and decentralized key management.

\subsection{Current Web3 Infrastructure Security Approaches}

Existing Web3 infrastructure security solutions primarily focus on consensus mechanisms and cryptographic protocols at the application layer. Projects like Secret Network~\cite{secret2020} and Oasis Network~\cite{oasis2020} have attempted to integrate confidential computing capabilities into blockchain networks. However, these solutions often rely on specific TEE implementations, introducing vendor dependencies and centralization risks.

Recent research has explored various approaches to enhance TEE security in Web3 contexts:

\begin{itemize}
\item Hybrid protocols combining TEE attestation with blockchain-based verification~\cite{zhang2019ekiden}
\item Multi-party computation protocols for distributed key management~\cite{boneh2018threshold}
\item Verifiable computation techniques for TEE program verification~\cite{wahby2018doubly}
\end{itemize}

While these approaches address specific aspects of TEE security, they fail to provide a comprehensive solution that maintains both security guarantees and Web3 principles of decentralization and trustlessness. Most existing solutions focus on individual components rather than providing an integrated framework.

\subsection{Gaps in Current Research}

Our analysis of existing literature reveals several critical gaps that dstack addresses:

\begin{enumerate}
\item \textbf{Limited TEE Portability}: Current solutions bind applications to specific hardware implementations, preventing migration and creating vendor lock-in
\item \textbf{Insufficient Blockchain Integration}: TEE security mechanisms are poorly integrated with blockchain-based governance, limiting transparency and auditability
\item \textbf{Incomplete Key Management}: Existing solutions lack comprehensive frameworks for decentralized attestation and key management that support key rotation and recovery
\item \textbf{Inadequate Failure Recovery}: Current systems provide limited consideration of failure modes and recovery mechanisms in TEE-based systems, violating the ``Assume Breach'' principle
\end{enumerate}

These gaps highlight the need for a comprehensive framework like dstack that systematically addresses the challenges of deploying confidential computing in Web3 environments while maintaining both security and decentralization guarantees.

\section{System Design}
\label{sec:design}

\begin{figure}[ht]
    \centering
    \includegraphics[width=\textwidth]{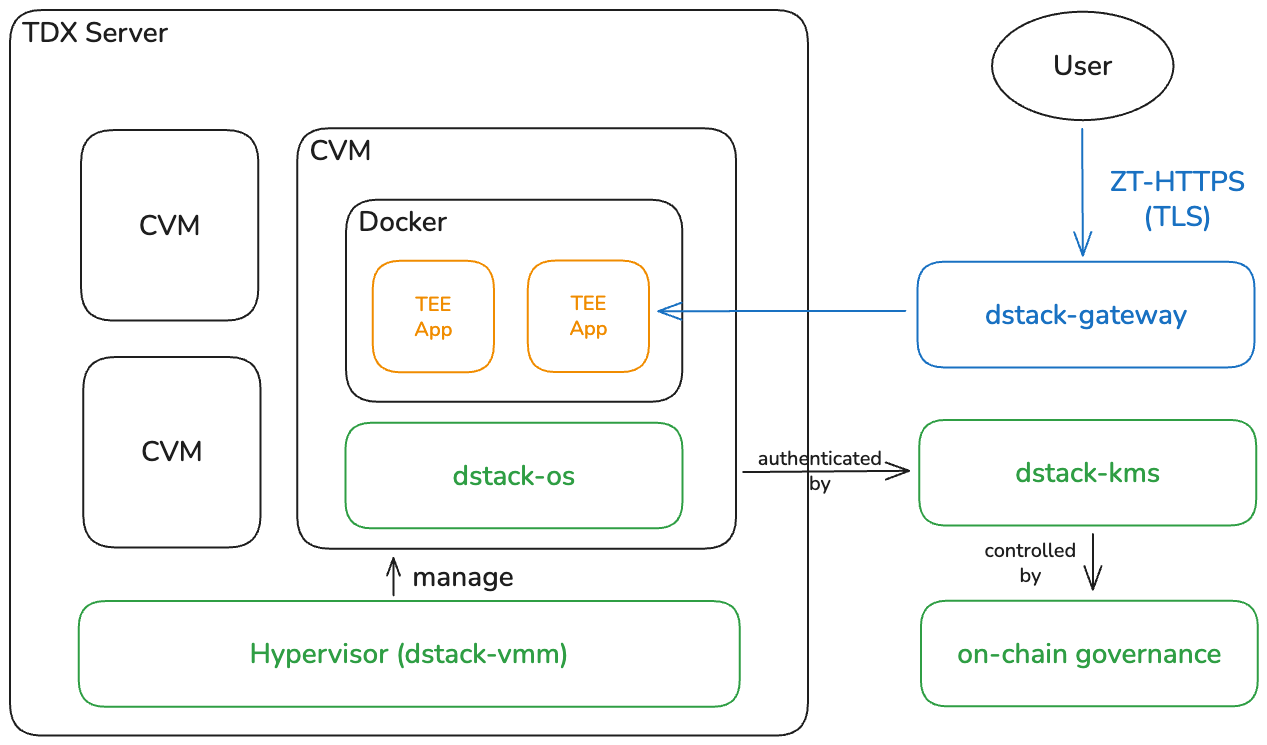}
    \caption{\small The architecture of dstack, including dstack-OS and dstack-KMS. Dstack-OS provides a hardware abstraction layer with a minimized operating system image, while dstack-KMS replaces hardware-based data encryption keys with a blockchain-controlled key management service.}
    \label{fig:system-design}
\end{figure}

\subsection{Portable Confidential Containers}

Containerization has transformed cloud-native application deployment by providing unparalleled portability and scalability through effective data separation. These capabilities are particularly critical for Trusted Execution Environment (TEE) applications, as they directly address the inherent limitations outlined in Section~\ref{sec:introduction}:

\begin{itemize}
    \item \textbf{Security Reliability}: Containerization enables applications to be transferred from TEE hardware that is known to be vulnerable (e.g., following responsible disclosure of a vulnerability) to secure instances, preserving code integrity. Our secure data migration mechanism, which ensures both forward and backward secrecy, can limit privacy leakage risks.
    \item \textbf{Censorship Resistance}: The architecture's inherent scalability in supporting multi-instance deployments of a single TEE application provides robust defense against Denial-of-Service attacks. Additionally, by decoupling data storage from execution environments, established data availability solutions can be readily integrated to ensure persistent access to application data.
    \item \textbf{Vendor Independence}: Our hardware abstraction layer enables confidential containers to operate across diverse TEE hardware platforms without modification, eliminating vendor lock-in while facilitating rapid adoption of emerging TEE technologies as they become available.
\end{itemize}

However, implementing truly portable confidential containers presents substantial technical challenges that differentiate them from conventional containerization. Two fundamental obstacles must be overcome:

First, existing TEE implementations generate data sealing keys derived from hardware-bound root keys unique to each TEE instance. This architectural design creates an inherent barrier to portability—data encrypted by one TEE instance cannot be decrypted by any other, even when running identical application code. Prior research has largely circumvented this limitation by restricting TEEs to stateless applications, but this approach severely constrains their utility in complex real-world scenarios.

Second, TEE implementations from different vendors impose disparate specifications for deployable programs. This fragmentation forces developers to create and maintain multiple artifacts across platforms, each requiring separate (and expensive) code reviews to verify functional equivalence—a process that undermines the trust guarantees central to Web3 applications.

To address these challenges comprehensively, we introduce two complementary components: dstack-KMS and dstack-OS.

\subsubsection{Dstack-KMS: Blockchain-Controlled Secret Derivation}
\label{sec:kms}

Dstack-KMS is a key derivation service that fundamentally transforms how encryption keys are managed in confidential computing environments. Its primary function is to generate a unique and stable secret (the \textit{Root Key}) for each application based on its code and configurations. This root key serves as the foundation for deriving additional application-specific secrets used for data encryption and verifiable random number generation.

Unlike hardware-based encryption approaches, dstack-KMS deliberately decouples key generation from specific TEE hardware instances. This architectural decision enables the critical capability of \textbf{encrypted data migration between different TEE instances}, with decryption possible after proper authorization. This feature is essential for achieving true censorship resistance.

Given its position as the root of trust for the entire system, the verifiability and availability of dstack-KMS are paramount. Unlike previous approaches~\cite{zhang2019ekiden,secret2020} that assume TEE infallibility, our threat model explicitly acknowledges that \textbf{TEEs can be compromised}. To address this reality, dstack-KMS implements comprehensive key rotation capabilities that provide both backward and forward secrecy for application data. This ensures that even if a specific TEE hardware instance is compromised, applications can be migrated with limited data exposure.

\begin{figure}[ht]
    \centering
    \includegraphics[width=0.6\textwidth]{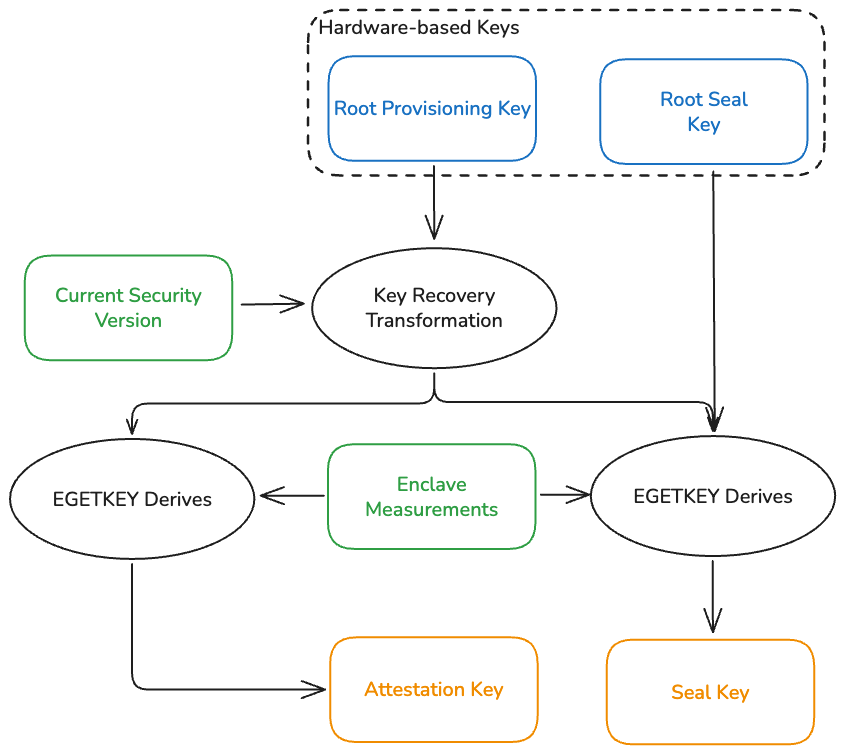}
    \caption{\small The key derivation hierarchy of Intel SGX illustrates how conventional TEE implementations bind encryption keys to hardware.}
    \label{fig:sgx-key-hierarchy}
\end{figure}

To understand the limitations of conventional approaches, Figure~\ref{fig:sgx-key-hierarchy} illustrates the key derivation hierarchy in Intel SGX. This design uses a hardware-bound key to both identify the TEE instance (for Remote Attestation) and generate encryption keys for data sealing. While this approach effectively establishes hardware identity, it fundamentally undermines censorship resistance:

\begin{itemize}
    \item The root key cannot be updated, meaning a compromised TEE with a leaked root key remains permanently vulnerable even after patching the underlying vulnerability.
    \item Encrypted data becomes intrinsically bound to specific TEE hardware instances, creating single points of failure that contradict Web3's censorship resistance requirements.
\end{itemize}

Our dstack-KMS addresses these limitations by delegating secret generation to an independent service that ensures encrypted data remains portable across TEE instances.

\begin{figure}[ht]
    \centering
    \includegraphics[width=0.7\textwidth]{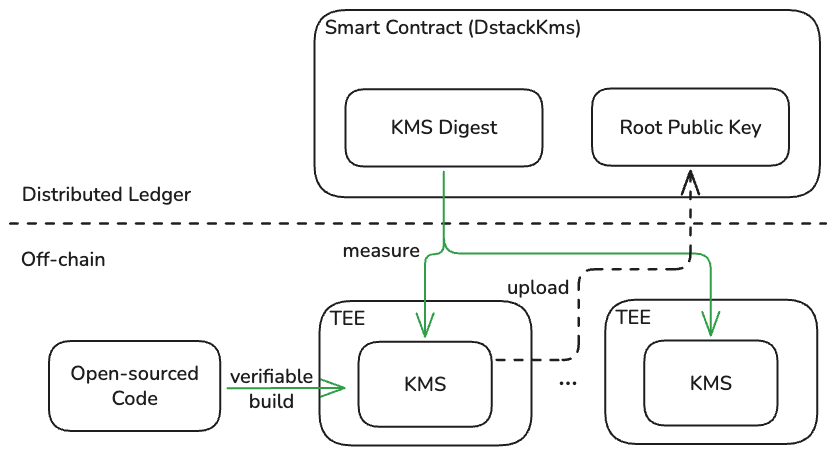}
    \caption{\small The architecture of dstack-KMS combines on-chain governance through smart contracts with an off-chain P2P network of secret derivation service nodes. This design ensures both verifiability and resilience against compromise.}
    \label{fig:kms}
\end{figure}

As illustrated in Figure~\ref{fig:kms}, dstack-KMS comprises an on-chain governance smart contract coupled with an off-chain peer-to-peer network where the actual secret derivation service operates.
To ensure comprehensive verifiability of the dstack-KMS program's integrity and correctness, we implement a multi-stage verification process:

\begin{enumerate}
    \item The dstack-KMS codebase is fully open-sourced, enabling thorough code and security reviews to verify key management logic and ensure the absence of backdoors.
    \item Executable binaries are produced through reproducible build processes, allowing verifiers to confirm that the runtime code matches the reviewed source code.
    \item The cryptographic digest of valid executables is published through the governance smart contract, creating an immutable reference for verification.
    \item Each service node operates within its own TEE instance. Critically, we rely on the TEE solely for measuring the executable and ensuring it matches the on-chain digest—not for the fundamental security of the key management process.
\end{enumerate}

This verification framework ensures that dstack-KMS remains trustworthy to any third-party verifier while avoiding dependency on the absolute security of any single TEE instance.

The core logic of each dstack-KMS service node is intentionally minimalist, focusing exclusively on the generation and secure distribution of the root key, along with deriving encryption keys for TEE applications. This simplicity enables the implementation of Multi-Party Computation (MPC)-based threshold key generation and derivation without performance overhead concerns. We have developed two complementary implementations:

\begin{itemize}
    \item \textbf{Simple Duplication}: The first node in the P2P network generates a cryptographically secure random number as the root key, then shares it with other nodes after verifying their attestation reports. All nodes maintain identical copies of the root key and derive consistent encryption keys for applications. This approach maximizes availability—as long as at least one node remains operational, the root key can be recovered.
    \item \textbf{MPC-Based Key Generation}: While simple duplication provides robust availability, it creates vulnerability to single-node compromises. Our MPC implementation uses Shamir's Secret Sharing scheme~\cite{shamir1979share} to distribute the root key across multiple nodes, ensuring that compromising any individual node (or even up to $t-1$ nodes, where $t$ is the configured threshold) does not expose the root key. This approach also enables advanced features like key rotation without requiring application reconfiguration.
\end{itemize}

\paragraph{Key Derivation}
Dstack-KMS implements a comprehensive key derivation framework to generate application-specific keys from the root key:

\begin{itemize}
    \item \textbf{Application CA Key}: Derived from the root CA key using the application's unique identifier (calculated as the hash of the application's code and configuration) following HKDF principles~\cite{krawczyk2010cryptographic}.
    \item \textbf{Disk Encryption Key}: Derived using a combination of the application identifier and instance identifier, enabling secure storage with portability.
    \item \textbf{Environment Encryption Key}: Derived using the application identifier alone, allowing secure environment variable management across instances.
    \item \textbf{ECDSA Key}: Derived from the root ECDSA key, providing applications with consistent cryptographic identity for signing operations.
\end{itemize}

Each derived key maintains cryptographic isolation while ensuring consistency across different TEE instances running the same application.

\paragraph{Key Rotation}
A critical security feature of dstack-KMS is its support for key rotation, which significantly limits the exposure window during potential attacks while ensuring both forward and backward secrecy for sealed data. Key rotation procedures are initiated exclusively through the governance smart contract, ensuring transparency and auditability.
We discuss two distinct rotation mechanisms:

\begin{itemize}
    \item \textbf{Root Key Share Rotation}: Our MPC-based implementation enables rotation of individual key shares without modifying the root key itself. This process enhances root key security by regularly updating the distribution mechanism while remaining transparent to applications, as all derived application keys remain unchanged.
    \item \textbf{Root Key Rotation}: In scenarios where the root key may have been compromised, we support complete root key rotation. This process generates an entirely new root key and implements a controlled handover period during which both old and new keys remain valid. This transition window allows applications to re-encrypt their data with keys derived from the new root key before the old root key is permanently destroyed, minimizing service disruption while enhancing security.
\end{itemize}

Through these mechanisms, dstack-KMS provides the foundation for truly portable confidential containers by decoupling encryption from hardware dependencies while maintaining robust security guarantees.

\subsubsection{Dstack-OS: Hardware Abstraction Layer}

Dstack-OS provides a comprehensive hardware abstraction layer with a minimized operating system image that bridges the gap between application containers and VM-level TEE implementations.
This layer enables applications to deploy across diverse TEE environments without code modifications while maintaining security integrity.
As a shared component across all TEE applications, dstack-OS must be intrinsically secure and verifiable, with no administrative backdoors that could compromise program code or user data.

VM-level TEE solutions typically require developers to provide complete system images encompassing bootloader, kernel, operating system, and application components.
This requirement not only increases development complexity but also introduces potential security vulnerabilities through misconfigured system components. A secure and properly configured operating system environment fundamentally determines the security baseline for all TEE applications built upon it.

\begin{figure}[ht]
    \centering
    \includegraphics[width=0.6\textwidth]{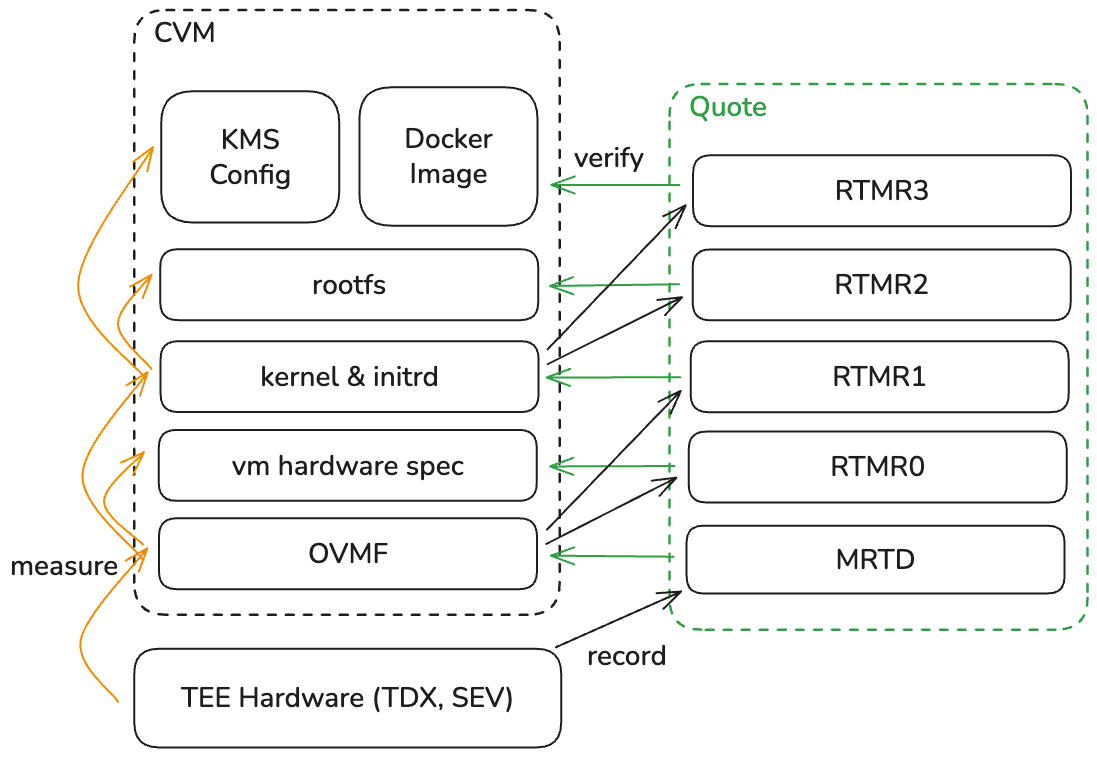}
    \caption{\small The architecture of dstack-OS establishes a secure and verifiable startup chain from TEE hardware through to user applications.}
    \label{fig:dos}
\end{figure}

Figure~\ref{fig:dos} illustrates how dstack-OS establishes a secure and verifiable startup chain from the TEE hardware foundation through to user-provided container images.
Our modular design ensures that the majority of the system can be reused across different TEE hardware implementations, substantially reducing the effort required for code review and security auditing. Combined with reproducible build processes, this approach enables comprehensive verification of dstack-OS with minimal overhead while supporting multiple TEE hardware platforms.

Using Intel TDX as an exemplar (with similar procedures applicable to other TEE implementations like AMD SEV), dstack-OS consists of the following components, presented in boot sequence order.
During this process, each component measures the next one and records these measurements in hardware registers, which are ultimately incorporated into the TEE attestation report:

\begin{itemize}
    \item \textbf{Hypervisor}: When application deployment initiates, the underlying hypervisor (the TDX module) loads the Open Virtual Machine Firmware (OVMF), enabling UEFI support for confidential virtual machines. The TDX module measures the OVMF code and records its cryptographic digest in the MRTD register, with the hypervisor's integrity guaranteed by the TEE hardware itself.
    \item \textbf{Open Virtual Machine Firmware (OVMF)}: The OVMF configures virtual machine hardware specifications (CPU, memory, and device configurations) and records the configuration measurement to the RTMR0 register. It then loads the Linux kernel and records the kernel image measurement to the RTMR1 register.
    \item \textbf{Kernel}: The kernel first loads an initial ramdisk (initrd) — a temporary in-memory filesystem containing minimal command-line tools and disk encryption libraries. The initrd mounts the root filesystem and records this measurement in the RTMR2 register.
    \item \textbf{Root Filesystem}: The root filesystem (RootFs) is a read-only environment containing essential system libraries and tools. It manages the entire lifecycle of deployed applications, measuring application images along with their associated key management system and storing these measurements in the RTMR3 register.
          The RootFs serves two primary functions: (1) providing the runtime environment for user-level container images, and (2) managing application data encryption.
          It interfaces with dstack-KMS to acquire encryption secrets and implements Linux Unified Key Setup (LUKS) to encrypt application data in isolated volumes. Data integrity is ensured through dm-verity, a Merkle-tree-based block-level verification tool.
          Finally, the RootFs configures network ingress for the application and initiates execution.
\end{itemize}

Rather than adapting existing system images, we built dstack-OS from scratch to ensure only essential components are included in both the initrd and RootFs.
A comprehensive component inventory is available in our implementation repository for transparency and review. Generally, dstack-OS includes only basic utilities (busybox), filesystem support, and container runtime components. By implementing Docker~\cite{merkel2014docker} support while delegating orchestration to the application level, we maintain a minimal base image footprint.

To ensure comprehensive verifiability, dstack-OS is fully open-sourced and supports reproducible builds~\cite{lamb2017reproducible}, allowing anyone to review the code and generate identical artifacts for verification. This approach provides significant benefits for applications requiring security reviews, as evaluators need only review the application code rather than the entire system image.

\paragraph{Data Backup and Defense against Rollback Attacks}
A critical challenge in confidential computing is ensuring data durability and integrity, particularly against rollback attacks where adversaries attempt to revert applications to previous states to bypass security checks or reuse expired secrets.

Our architecture provides robust secure data backup support and implements robust anti-rollback mechanisms that application developers can integrate:

\begin{itemize}
    \item \textbf{Secure Backup}: Application data is encrypted using keys derived from dstack-KMS and can be backed up to external storage providers. Since encryption keys are not bound to specific TEE instances, data can be restored on any authorized TEE, supporting true portability and disaster recovery.
    \item \textbf{Monotonic Counters}: To prevent rollback attacks, developers can implement monotonic counters that increment with each critical data update, incorporating the counter value into the encrypted data. During restoration, applications verify that the counter value equals or exceeds the last known value, preventing replay of outdated snapshots.
\end{itemize}

\noindent By combining these techniques, dstack ensures both the recoverability and integrity of confidential data across portable container migrations and backup scenarios.

Together, dstack-KMS and dstack-OS provide the foundational infrastructure for truly portable confidential containers that maintain security and integrity across heterogeneous TEE environments while eliminating vendor lock-in and supporting comprehensive verifiability.

\subsection{Decentralized Code Management}

A core principle of Zero Trust platforms is the enforcement of ``Code is Law'', ensuring that application code operates as intended without unexpected modifications.
While TEE hardware provides code integrity guarantees during execution, it cannot independently prevent administrators from deploying malicious code or making unauthorized modifications.
Conversely, smart contracts have established robust methodologies for implementing programmable governance over code lifecycle management — such as multi-signature requirements that prevent single-actor manipulation.

Our decentralized code management framework bridges these two worlds by placing TEE application governance under smart contract control, creating a transparent and auditable system for application deployment and updates that aligns with Web3 principles.

\begin{figure}[ht]
    \centering
    \includegraphics[width=0.8\textwidth]{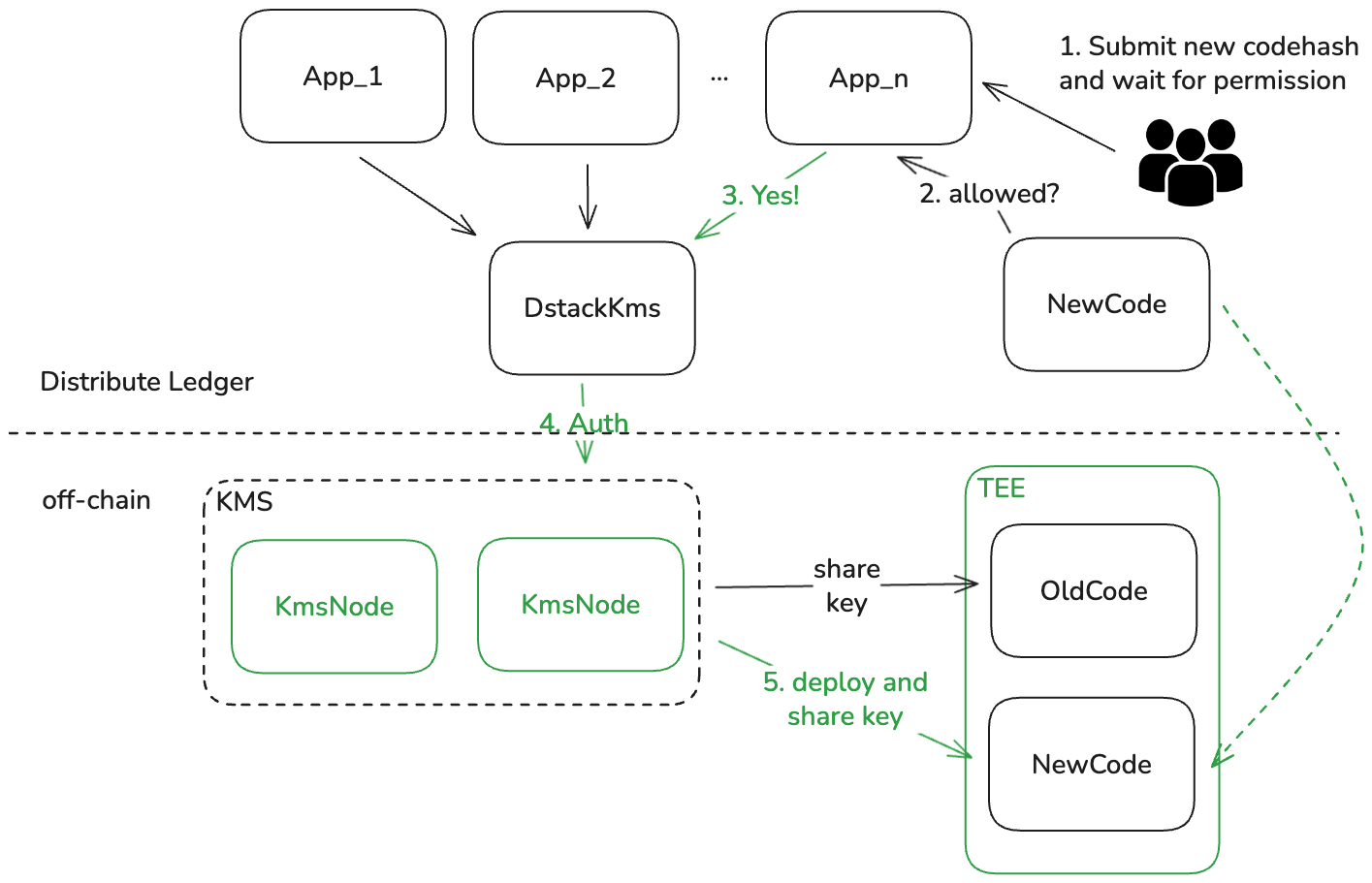}
    \caption{\small The architecture of decentralized code management integrates on-chain governance contracts with off-chain TEE through dstack-KMS.}
    \label{fig:code-govern}
\end{figure}

As illustrated in Figure~\ref{fig:code-govern}, our architecture consists of two complementary components: on-chain governance smart contracts and off-chain TEE execution environments connected through dstack-KMS.
This design establishes the on-chain contracts as the definitive root of trust for the entire system. Every operation — from initial deployment to code upgrades — must first be initiated and authorized through the governance contracts before off-chain TEE instances can execute these operations.

The governance framework implements a two-tier structure:

\begin{itemize}
    \item \textbf{KmsAuth Contract}: Serves as the global authority controlling dstack-KMS operations. This contract maintains the registry of authorized applications and their governance parameters, enforcing system-wide policies for TEE deployment. By controlling whether dstack-KMS shares application secrets with specific TEE instances, the KmsAuth contract effectively determines whether applications can decrypt their data, thus controlling their complete lifecycle.
    \item \textbf{AppAuth Contracts}: Individual governance contracts deployed for each application that define application-specific management rules. These contracts specify permissible code versions (through cryptographic hashes), authorized TEE instance identities, and upgrade approval requirements. This modular approach enables customized governance models ranging from traditional multi-signature schemes to complex DAO voting mechanisms, allowing each application to implement governance appropriate to its requirements.
\end{itemize}

The enforcement mechanism operates through dstack-KMS, which will only provide application secrets to TEE instances running code versions explicitly authorized by the governance contracts. This creates a cryptographically enforced governance system where unauthorized code versions cannot access application data, regardless of administrative privileges.

To illustrate this mechanism, consider a code upgrade process for an application using a multi-signature AppAuth contract in Figure~\ref{fig:code-govern}:

\begin{enumerate}
    \item Developers publish the new code version and submit its cryptographic hash to the AppAuth contract.
    \item The contract initiates the approval process, requiring signatures from designated key holders (e.g., core developers, security auditors, and community representatives).
    \item Each signature is recorded on-chain, creating an immutable audit trail of the approval process.
    \item Once the required signature threshold is reached, the AppAuth contract updates its registry of authorized code versions.
    \item The KmsAuth contract, monitoring the AppAuth contract, updates its authorization records accordingly.
    \item dstack-KMS, which continuously synchronizes with the KmsAuth contract, begins providing application secrets to TEE instances running the newly approved code version.
    \item TEE instances can now deploy the new code with full access to encrypted application data.
\end{enumerate}

\noindent This process ensures that every code change follows predefined governance rules with complete transparency. The on-chain approval records create an immutable audit trail that allows users to verify the entire history of application changes, while the cryptographic enforcement through dstack-KMS guarantees that only properly authorized code can access application data.

By combining the code integrity guarantees of TEEs with the transparent governance capabilities of smart contracts, our decentralized code management framework ensures that applications truly embody the "Code is Law" principle. Every aspect of the application lifecycle—from initial deployment to version updates and potential retirement—follows predefined, transparent rules that cannot be circumvented by any single entity, including the platform operators themselves.

\subsection{Verifiable Domain Management}

A complete Zero Trust platform must enable seamless verification for users across both Web3 and traditional Web2 environments.
Our Verifiable Domain Management system completes the chain of trust by allowing standard Web browsers to cryptographically verify TEE applications without requiring any client-side modifications or specialized knowledge.

This capability is critical for mainstream adoption, as it allows users to interact with confidential applications using familiar tools while maintaining the same security guarantees provided by the underlying TEE infrastructure. We implement this through two complementary components—dstack-Ingress and dstack-Gateway—both leveraging our novel Zero Trust TLS (ZT-TLS) protocol.

\begin{figure}[ht]
    \centering
    \includegraphics[width=0.8\textwidth]{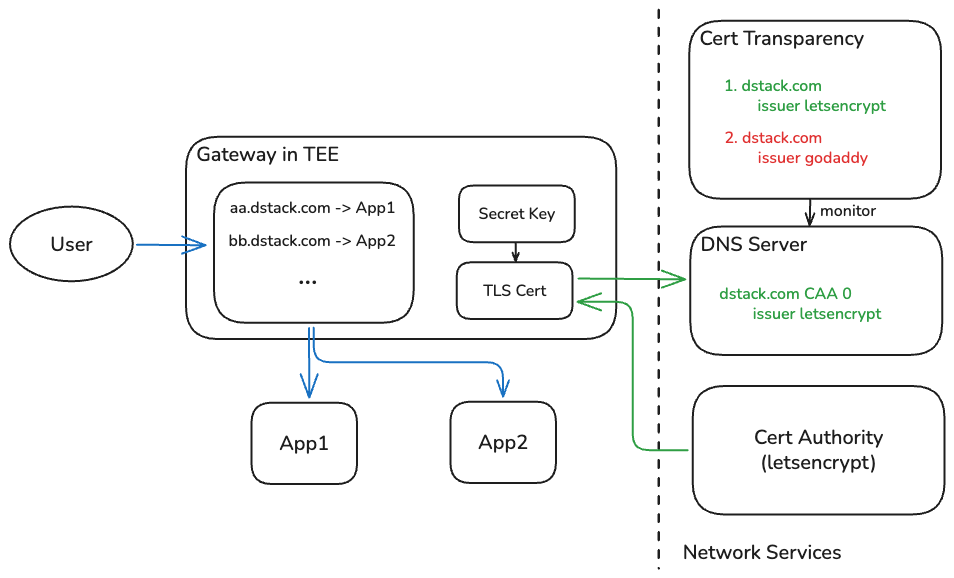}
    \caption{\small The Zero Trust TLS architecture establishes a cryptographically verifiable connection between standard web browsers and TEE applications through blockchain-anchored certificate management.}
    \label{fig:tee-domain}
\end{figure}

The fundamental innovation in ZT-TLS is establishing cryptographic proof that a specific TLS certificate (and therefore a domain) is exclusively controlled by a verified TEE application. This creates a verification chain extending from the blockchain to standard HTTPS connections, enabling users to confirm application authenticity through familiar security indicators in their browsers.

\subsubsection{Zero Trust TLS Protocol}
\label{sec:zt-tls}

The ZT-TLS protocol addresses a critical challenge: how to cryptographically bind a domain to a specific TEE application in a way that prevents unauthorized certificate issuance or domain hijacking. Our solution integrates three key components:

\begin{itemize}
    \item \textbf{TEE-Generated Certificates}: The TEE application (specifically dstack-Gateway) generates its TLS certificate using application secrets derived from dstack-KMS. This ensures the private key never exists outside the TEE environment and creates a cryptographic link between the on-chain governance and the TLS certificate.
    \item \textbf{Certificate Authority Authorization (CAA)}: We leverage DNS CAA records to restrict which certificate authorities can issue certificates for managed domains. By configuring CAA records to only authorize certificates from specific issuers (which verify TEE attestation), we prevent unauthorized certificate issuance through alternative validation methods.
    \item \textbf{Certificate Transparency (CT) Monitoring}: To protect against unauthorized DNS record modifications, we implement continuous monitoring of Certificate Transparency logs~\cite{laurie2013certificate}. This allows detection of any unauthorized certificates and provides an immutable audit trail of certificate issuance history.
\end{itemize}

\subsubsection{Implementation Components}

We provide two complementary implementations to accommodate different application requirements:

\begin{itemize}
    \item \textbf{dstack-Gateway}: A fully managed reverse proxy running within a TEE that provides immediate domain verification with zero application code changes. Applications register with dstack-Gateway through a simple API call, during which dstack-Gateway verifies the application's remote attestation report. Once verified, applications receive a subdomain under a pre-registered wildcard domain (e.g., \texttt{app-id.dstack.com}), with all TLS termination and certificate management handled automatically.
    \item \textbf{dstack-Ingress}: A more flexible solution for applications requiring custom domain names. Dstack-Ingress provides libraries and services that enable applications to manage their own TLS certificates while maintaining the same verification guarantees. This approach requires minimal integration work while supporting application-specific domain requirements.
\end{itemize}

\noindent Both implementations ensure that certificates are cryptographically bound to verified TEE environments, with the verification chain extending from the blockchain governance contracts through to the TLS certificates presented to users' browsers.

\subsubsection{Security Guarantees and Protections}

Our Verifiable Domain Management system provides several critical security guarantees:

\begin{itemize}
    \item \textbf{Certificate Provenance Verification}: Users can verify that the TLS certificate presented by an application was issued to a legitimate TEE environment by checking the certificate against on-chain records.
    \item \textbf{Domain Binding Integrity}: The combination of CAA records and Certificate Transparency monitoring ensures that domains remain bound to their designated TEE applications, preventing unauthorized certificate issuance.
    \item \textbf{Tamper-Evident Certificate Changes}: Any unauthorized attempts to modify domain configurations or issue alternative certificates are detected through Certificate Transparency monitoring, with alerts propagated to clients.
\end{itemize}

While we cannot technically prevent all attacks against centralized DNS infrastructure, our multi-layered approach creates a tamper-evident system where unauthorized changes are quickly detected and rejected by client verification processes. This approach aligns with the "Assume Breach" principle by implementing defense-in-depth strategies that maintain security even when individual components are compromised.

By extending the verification chain from blockchain governance through TEE attestation to standard TLS certificates, our Verifiable Domain Management system enables seamless integration of Zero Trust applications into existing web infrastructure. This creates a practical path for mainstream adoption without requiring specialized knowledge or tools from end users, while maintaining the comprehensive security guarantees that define true Zero Trust platforms.

\section{Verifying TEE Applications: A Practical Walkthrough}

To demonstrate how dstack fulfills the Zero Trust principles and addresses the challenges outlined in Section~\ref{sec:introduction}, we present a practical end-to-end verification scenario. This walkthrough illustrates how the architectural components described in Section~\ref{sec:design} work together to provide a comprehensive chain of trust for real-world Web3 applications.

Traditional Zero Trust programs like smart contracts achieve verifiability through transparent source code and inputs. However, extending similar verifiability to real-world applications presents unique challenges, particularly in maintaining privacy and compatibility with existing development practices. Dstack addresses these challenges through a comprehensive verification framework that requires minimal code changes.

This section demonstrates dstack's complete chain of trust through a practical example: verifying that a cryptographic key was genuinely generated within a TEE environment. This scenario is especially relevant for Web3 and AI agent applications where cryptographic keys establish application identity and authority.

\subsection{Threat Model}

Our verification framework assumes the correct implementation and integrity of core dstack components (dstack-OS, dstack-KMS, dstack-Gateway, and governance contracts). While these components' security guarantees are detailed in later sections, we explicitly consider developers and administrators as potential adversaries.

Potential attack vectors include:

\begin{itemize}
    \item \textbf{Application-Level Attacks}
          \begin{itemize}
              \item Deployment of unauthorized or malicious code through falsified Docker images
              \item Introduction of malicious dependencies through supply chain attacks
              \item Unauthorized application updates that bypass governance controls
          \end{itemize}

    \item \textbf{Infrastructure-Level Attacks}
          \begin{itemize}
              \item Tampering with dstack-OS to circumvent security controls
              \item Deployment of compromised key management services
              \item Operation of unpatched or vulnerable TEE firmware
              \item Execution of applications on non-TEE hardware with spoofed attestation
          \end{itemize}

    \item \textbf{Network-Level Attacks}
          \begin{itemize}
              \item DNS hijacking to redirect traffic to unauthorized applications
              \item Man-in-the-middle attacks on communication channels
          \end{itemize}
\end{itemize}

Direct attacks on TEE hardware security (e.g., exploitation of hardware vulnerabilities) are considered out of scope for this discussion and addressed separately.

\subsection{The Reproducibility Challenge}

A fundamental challenge in TEE attestation lies in bridging the gap between source code and the actual artifacts (binaries or container images) running inside enclaves. While TEE attestation can prove that a specific artifact runs within a secure environment, it cannot independently verify that the artifact accurately represents its claimed source code.

This limitation creates several security risks:

\begin{itemize}
    \item \textbf{Source Code Manipulation}: Developers could intentionally deploy artifacts that differ from publicly reviewed source code
    \item \textbf{Build Process Compromise}: Supply chain attacks could inject malicious code during artifact creation
    \item \textbf{Toolchain Vulnerabilities}: As demonstrated by the Ken Thompson Hack, even compiler toolchains could be compromised to inject malicious behavior
\end{itemize}

TEE quotes only attest to measurements (cryptographic hashes) of running artifacts. Without guarantees that these artifacts accurately represent reviewed source code, users cannot fully trust applications even when running in secure enclaves. This challenge affects all TEE applications, including dstack's core components.

\begin{figure}[ht]
    \centering
    \includegraphics[width=0.8\textwidth]{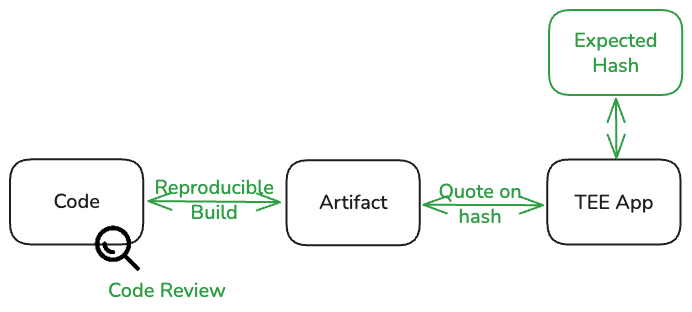}
    \caption{\small Comprehensive verification chain from source code to running TEE applications}
    \label{fig:reproducible-build}
\end{figure}

Establishing a complete verification chain requires three key elements:

\begin{itemize}
    \item \textbf{Source Code Verification}: Open-source code or trusted third-party review ensuring no backdoors exist
    \item \textbf{Build Process Integrity}: Reproducible builds allowing independent verification of artifacts
    \item \textbf{Runtime Attestation}: TEE quotes proving that verified artifacts are actually running
\end{itemize}

Traditional reproducible builds, while ideal, often introduce significant complexity through specialized toolchains and build scripts. The following section introduces an alternative approach, \emph{In-CC Build}, that achieves similar guarantees with improved practicality.

\subsection{Dstack Component Verifiability}

Each core dstack component implements specific verifiability mechanisms anchored in transparent, immutable records. Table~\ref{tab:trust-base} summarizes these mechanisms:

\begin{table}[ht]
    \centering
    \begin{tabular}{|l|c|c|c|l|}
        \hline
        \textbf{Component} & \textbf{Source} & \textbf{Build Method} & \textbf{Registry} & \textbf{Verification}   \\
        \hline
        dstack-OS          & Open            & Reproducible          & KmsAuth           & OS Digest               \\
        \hline
        dstack-KMS         & Open            & In-CC                 & KmsAuth           & Image Hash, Root PubKey \\
        \hline
        dstack-Gateway     & Open            & In-CC                 & AppAuth           & Image Hash              \\
        \hline
    \end{tabular}
    \caption{\small Verifiability mechanisms and trust anchors for dstack components}
    \label{tab:trust-base}
\end{table}

All components maintain open-source codebases with comprehensive documentation. Dstack-OS, as the foundational component, undergoes third-party security audits and implements fully reproducible builds.

The governance smart contracts serve as authoritative registries, anchoring each component's trust base:

\begin{itemize}
    \item Dstack-OS digests are registered in KmsAuth and managed through multi-signature governance
    \item Dstack-KMS measurements and root public keys are recorded in KmsAuth
    \item Dstack-Gateway artifacts are verified through dedicated AppAuth contracts
\end{itemize}

\paragraph{In-ConfidentialContainer Build}
While reproducible builds represent the gold standard for artifact verification, they often prove impractical for complex applications. Dstack addresses this through \emph{In-CC Build}, leveraging the verified dstack-OS as a trusted foundation.
The In-CC Build process executes entirely within dstack-OS environment, encompassing code fetching, compilation, and artifact creation. By specifying precise source references (e.g., a snapshot of the codebase), the system produces verifiable artifacts with measurements attested in TEE quotes. This approach minimizes the trusted computing base while eliminating the complexity of traditional reproducible builds.

\subsection{End-to-End Verification Walkthrough}

From a user's perspective, verifying a TEE application in dstack requires validating a chain of measurements against authoritative registries, encompassing the user application, runtime data, dstack components, and network infrastructure.
This section walks through the verification process, demonstrating how each step builds upon the previous one to create a comprehensive chain of trust.

\begin{figure}[ht]
    \centering
    \includegraphics[width=\textwidth]{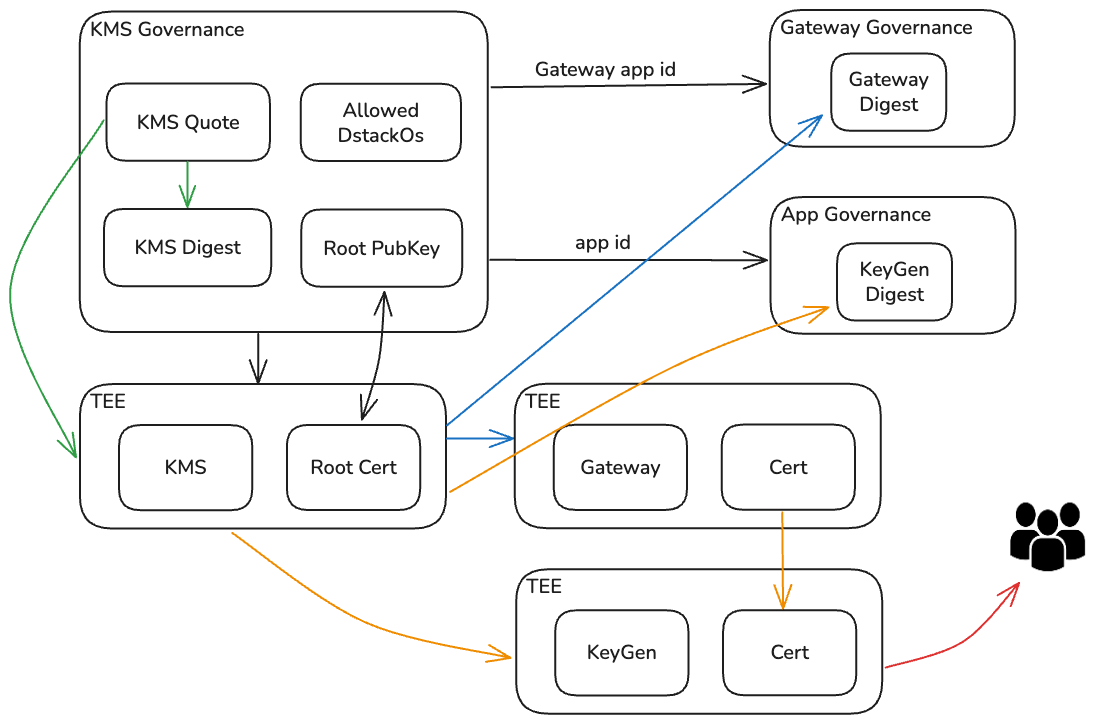}
    \caption{\small The verification chain from authoritative registries to runtime measurements}
    \label{fig:e2e-verification}
\end{figure}

As shown in Figure~\ref{fig:e2e-verification}, the verification chain begins with the KmsAuth smart contract, which serves as the root of trust for the entire system,
and concludes with the key generation application running in a TEE.
We explain the verification steps as follows.

First, the address of the dstack-KMS governance contract (i.e., KmsAuth) is publicly available. The KMS governance contract serves as the authoritative registry for (1) off-chain KMS nodes and (2) dedicated governance contracts (i.e., AppAuth) for TEE applications.
It is worth noting that the dstack-Gateway can be regarded as a special TEE application that attests and forwards requests to other user applications while maintaining consistent verification procedures.

The KmsAuth contract records both the measurements of off-chain KMS node programs (KMS digest, which is the hash of the docker compose file and the images used in In-CC Build) and the TDX Quote of the first KMS node, which acts as the primary key generator. The identities of subsequent KMS nodes must first be registered in KmsAuth and verified by the first KMS node before they can participate in key generation.
These data, together with the open-source dstack-KMS codebase, allow users to verify the integrity of the KMS nodes and their code correctness. Subsequently, the off-chain KMS nodes can be considered trusted extensions and used to verify Quotes from other TEE applications, which cannot be achieved by the on-chain governance contract alone.
The off-chain KMS nodes are responsible for generating and managing cryptographic keys (i.e., the root key) and the corresponding certificates (explained in Section~\ref{sec:zt-tls}), which are further used to authorize TEE applications after they are verified, thereby avoiding redundant verification.

Second, users can verify the validity of dstack-Gateway. By default, this is delegated to the off-chain KMS nodes during gateway registration, although users can always request the Quote of the gateway and check the digest of its docker compose file in its governance contract and the codebase.
After the KMS nodes verify the gateway, it will generate its application root key along with a certificate chain ending with the root certificate.

Next, users can access the TEE application through dstack-Gateway forwarding. In this case, users effectively delegate the verification of the key generation application to the gateway.
The deployment of the key generation application starts from the KmsAuth contract. The deployer first registers it and stores the digest (the hash of the docker compose file and the images) on the dedicated AppAuth contract.
Then, as in the previous step, the off-chain KMS nodes verify the Quote of the application after it is deployed to the TEE and ensure it matches the registered digest. Only after successful verification will the key be generated along with the certificate chain.
To be forwarded by the gateway, the application must also register with it. In this case, the gateway verifies the application's certificate chain and ensures it shares the same root certificate, indicating authorization by the same KMS.

Finally, the user can invoke the application and request proof of the key it generated. The application can generate a Quote on demand, with the generated public key included as custom data in the Quote.

\subsection{Security Analysis: Mitigating Threats}

Our key generation example demonstrates dstack's comprehensive approach to security across multiple attack surfaces. At the application level, the system prevents unauthorized code deployment through rigorous measurement verification against the AppAuth registry.
The In-CC Build process ensures all dependencies originate from verified sources, while any updates require new registry entries and complete re-verification.
It is important to note, however, that code audit and review processes remain essential for ensuring application code correctness, as the In-CC Build only guarantees the integrity of the build process.

Infrastructure security is maintained through multiple mechanisms. TDX Quotes verify dstack-OS integrity, preventing OS-level tampering. KMS updates require multi-signature governance approval, safeguarding against unauthorized changes to key management services. The Quote verification process includes firmware version validation, ensuring applications run on up-to-date TEE hardware. Hardware-backed TDX Quotes effectively prevent attestation spoofing attempts.

Network-level threats are mitigated through the certificate chain validation system. This prevents both DNS hijacking and man-in-the-middle attacks by ensuring authentic application identity and enabling end-to-end encryption with verified certificates. The combination of these security measures ensures that cryptographic keys maintain their integrity guarantees even when confronted with sophisticated attacks targeting different system layers.

\section{Discussion}

This section examines the broader implications of dstack, discusses current limitations, and outlines future research directions. We address key challenges that remain outside the scope of our current implementation while highlighting opportunities for further advancement.

\subsection{Security Implications and Threat Model}

While dstack significantly advances the security posture of confidential computing in Web3 environments, several important considerations merit discussion.

\subsubsection{TEE Hardware Vulnerabilities}

Dstack's security model assumes that TEE hardware provides basic isolation and attestation capabilities. However, the discovery of vulnerabilities in TEE implementations (such as Foreshadow, PlunderVolt, and Load Value Injection attacks) raises important questions about long-term security guarantees.

Our approach addresses this challenge through several mechanisms:

\begin{itemize}
\item \textbf{Defense in Depth}: Rather than relying solely on TEE hardware security, we implement multiple layers of protection through blockchain governance and key rotation
\item \textbf{Rapid Response}: Our key rotation capabilities enable quick response to disclosed vulnerabilities by migrating applications to patched hardware
\item \textbf{Distributed Trust}: The MPC-based key management reduces the impact of any single TEE compromise
\end{itemize}

However, detecting TEE exploitation remains a fundamental challenge. Unlike traditional software vulnerabilities, TEE compromises may be difficult to detect through conventional monitoring approaches.

\subsubsection{Detection of TEE Exploitation}

The detection of TEE exploitation represents one of the most challenging aspects of confidential computing security and remains largely out of scope for this paper. Current approaches are limited:

\begin{itemize}
\item \textbf{Statistical Analysis}: Monitoring for unusual patterns in application behavior or performance characteristics
\item \textbf{Honeypot Deployments}: Deploying decoy applications with known secrets to detect unauthorized access
\item \textbf{Cross-Validation}: Running multiple instances of critical computations and comparing results
\end{itemize}

Future work should explore more sophisticated detection mechanisms, potentially involving machine learning approaches to identify anomalous behavior patterns in TEE environments.

\subsection{Broader Impact}

Dstack's approach to combining confidential computing with decentralized governance has implications beyond Web3 applications:

\begin{itemize}
\item \textbf{Enterprise Computing}: Enabling confidential computing in multi-organization collaborations
\item \textbf{Government Services}: Supporting privacy-preserving public services with transparent governance
\item \textbf{Research Computing}: Enabling secure multi-party computation for sensitive research data
\end{itemize}

The principles established by dstack could inform the development of trustworthy computing systems across diverse domains requiring both confidentiality and verifiability.

\section{Conclusion}

This paper presented dstack, a comprehensive framework that transforms raw Trusted Execution Environment (TEE) technology into a true Zero Trust platform aligned with Web3 principles. We addressed critical limitations in current TEE implementations that prevent their effective deployment in Web3 contexts, specifically security reliability, censorship vulnerability, incomplete verifiability, and unrestricted application lifecycle control.

The convergence of confidential computing and Web3 represents a critical evolution in how we design and deploy trustworthy systems. Dstack demonstrates that it is possible to maintain the security guarantees of TEE technology while embracing the decentralized, trustless principles that define Web3.

Our work establishes a foundation for the next generation of applications that require both confidentiality and verifiability. By systematically addressing the limitations of current TEE implementations and providing practical solutions, dstack enables developers to build truly Zero Trust applications that serve the needs of an increasingly decentralized digital economy.

The principles and techniques developed in dstack extend beyond the immediate Web3 use case, offering a blueprint for trustworthy computing systems in any environment where confidentiality, verifiability, and decentralized governance are essential requirements. As both confidential computing and blockchain technologies continue to mature, frameworks like dstack will become increasingly important for realizing the full potential of secure, decentralized applications.

\bibliographystyle{alpha}
\bibliography{bibfile}

\newcommand{\etalchar}[1]{$^{#1}$}
\begin{thebibliography}{VBMW{\etalchar{+}}18}

\bibitem[Ama25]{awsnitro}
Amazon.
\newblock Aws nitro system.
\newblock \url{https://aws.amazon.com/ec2/nitro/}, 2025.

\bibitem[BGH{\etalchar{+}}18]{boneh2018threshold}
Dan Boneh, Craig Gentry, Shai Halevi, Jonathan Katz, and Mariana Raykova.
\newblock Threshold cryptosystems from threshold fully homomorphic encryption.
\newblock {\em Journal of Cryptology}, 31(2):565--596, 2018.

\bibitem[{Goo}25]{gcptee}
{Google}.
\newblock Google cloud confidential vm overview.
\newblock \url{https://cloud.google.com/confidential-computing/confidential-vm/docs/confidential-vm-overview}, 2025.

\bibitem[HSS{\etalchar{+}}18]{hunt2018chiron}
Tyler Hunt, Congzheng Song, Reza Shokri, Vitaly Shmatikov, and Emmett Witchel.
\newblock Chiron: Privacy-preserving machine learning as a service.
\newblock In {\em arXiv preprint arXiv:1803.05961}, 2018.

\bibitem[Kap16]{kaplan2016amd}
David Kaplan.
\newblock Amd memory encryption.
\newblock In {\em Linux Security Summit}, 2016.

\bibitem[KE10]{krawczyk2010cryptographic}
Hugo Krawczyk and Pasi Eronen.
\newblock Cryptographic extraction and key derivation: The hkdf scheme.
\newblock {\em RFC 5869}, 2010.

\bibitem[LGY{\etalchar{+}}19]{li2019intel}
Peng Li, Le~Guan, Xing Yu, Yi~Zhou, Lei Zhao, Robert~H Deng, Peng Liu, and Surya Nepal.
\newblock Intel trust domain extensions.
\newblock In {\em Proceedings of the 2019 ACM SIGSAC Conference on Computer and Communications Security}, pages 1--16, 2019.

\bibitem[LLK13]{laurie2013certificate}
Ben Laurie, Adam Langley, and Emilia Kasper.
\newblock Certificate transparency.
\newblock In {\em RFC 6962}, 2013.

\bibitem[LZ17]{lamb2017reproducible}
Chris Lamb and Stefano Zacchiroli.
\newblock Reproducible builds: Increasing the integrity of software supply chains.
\newblock In {\em IEEE Software}, volume~34, pages 64--70, 2017.

\bibitem[Mer14]{merkel2014docker}
Dirk Merkel.
\newblock Docker: lightweight linux containers for consistent development and deployment.
\newblock {\em Linux journal}, 2014(239):2, 2014.

\bibitem[Mic25]{azuretee}
Microsoft.
\newblock Azure confidential computing.
\newblock \url{https://learn.microsoft.com/en-us/azure/confidential-computing/}, 2025.

\bibitem[MOG{\etalchar{+}}20]{murdock2020plundervolt}
Kit Murdock, David Oswald, Flavio~D Garcia, Jo~Van~Bulck, Daniel Gruss, and Frank Piessens.
\newblock Plundervolt: Software-based fault injection attacks against intel sgx.
\newblock In {\em 2020 IEEE Symposium on Security and Privacy (SP)}, pages 1466--1482, 2020.

\bibitem[{Oas}20]{oasis2020}
{Oasis Protocol Foundation}.
\newblock Oasis network: A privacy-first cloud computing platform on blockchain.
\newblock {\em Whitepaper}, 2020.

\bibitem[{Sec}20]{secret2020}
{Secret Network}.
\newblock Secret network: Privacy-preserving smart contracts.
\newblock {\em Whitepaper}, 2020.

\bibitem[Sha79]{shamir1979share}
Adi Shamir.
\newblock How to share a secret.
\newblock {\em Communications of the ACM}, 22(11):612--613, 1979.

\bibitem[TB19]{tramer2018slalom}
Florian Tramer and Dan Boneh.
\newblock Slalom: Fast, verifiable and private execution of neural networks in trusted hardware.
\newblock In {\em International Conference on Learning Representations}, 2019.

\bibitem[VBMS{\etalchar{+}}20]{van2020lvi}
Jo~Van~Bulck, Daniel Moghimi, Michael Schwarz, Moritz Lipp, Marina Minkin, Daniel Genkin, Yuval Yarom, Berk Sunar, Daniel Gruss, and Frank Piessens.
\newblock Lvi: Hijacking transient execution through microarchitectural load value injection.
\newblock In {\em 2020 IEEE Symposium on Security and Privacy (SP)}, pages 54--72, 2020.

\bibitem[VBMW{\etalchar{+}}18]{van2018foreshadow}
Jo~Van~Bulck, Marina Minkin, Ofir Weisse, Daniel Genkin, Baris Kasikci, Frank Piessens, Mark Silberstein, Thomas~F Wenisch, Yuval Yarom, and Raoul Strackx.
\newblock Foreshadow: Extracting the keys to the intel sgx kingdom with transient out-of-order execution.
\newblock In {\em 27th USENIX Security Symposium (USENIX Security 18)}, pages 991--1008, 2018.

\bibitem[WTS{\etalchar{+}}18]{wahby2018doubly}
Riad~S Wahby, Ioanna Tzialla, Abhi Shelat, Justin Thaler, and Michael Walfish.
\newblock Doubly-efficient zksnarks without trusted setup.
\newblock In {\em 2018 IEEE Symposium on Security and Privacy (SP)}, pages 926--943, 2018.

\bibitem[ZZZ{\etalchar{+}}19]{zhang2019ekiden}
Chao Zhang, Mingyu Zhu, Fan Zhang, Haoxian Chen, Wenhao Qiu, Wei Chen, Cheng Han, and Dawn Song.
\newblock Ekiden: A platform for confidentiality-preserving, trustworthy, and performant smart contracts.
\newblock In {\em Proceedings of the 2019 IEEE European Symposium on Security and Privacy (EuroS\&P)}, pages 185--200, 2019.

\end{thebibliography}

\end{document}